\documentclass[sigconf]{acmart}
\usepackage[utf8]{inputenc} 
\pdfoutput=1
\setcopyright{none}
\renewcommand\footnotetextcopyrightpermission[1]{}

\usepackage{booktabs, multirow} 
\usepackage{soul}

\begin{document}

\title{ProFuzzBench: A Benchmark for Stateful Protocol Fuzzing}


\author{Roberto Natella}
\email{roberto.natella@unina.it}
\orcid{0000-0003-1084-4824}
\affiliation{%
  \institution{Università degli Studi di Napoli Federico II, Italy}
}

\author{Van-Thuan Pham}
\orcid{0000-0002-9871-3695}
\email{thuan.pham@unimelb.edu.au}
\affiliation{%
  \institution{University of Melbourne, Australia}
}


\begin{abstract}
We present a new benchmark (\emph{ProFuzzBench}) for stateful fuzzing of network protocols. The benchmark includes a suite of representative open-source network servers for popular protocols, and tools to automate experimentation. We discuss challenges and potential directions for future research based on this benchmark.
\end{abstract}



\keywords{Fuzzing; Benchmarking; Network Protocols}

\maketitle
\pagestyle{plain}

\section{Introduction}

Protocols are pervasive in computer systems, as they enable communication among parties over a local network or the internet, such as, for instant messaging, email, file serving and sharing, multimedia streaming, and several other applications. 
Therefore, protocol \emph{implementations} are an appealing target for malicious actors, as a vulnerability in an implementation may be remotely exploited. 
However, it is challenging to test the security of protocol implementations, since protocols are most often \emph{stateful} in nature. This means that, compared to stateless programs, the input space for testing is not limited to the format of individual messages (which can be very large by its own), but is further enlarged by the potential combinations of several messages. Therefore, protocol security testing is in need of techniques that can efficiently test such input space, by taking into account the protocol states.

Much of recent research on security testing has been on \emph{fuzzing} \cite{fuzz, fuzzsurvey}, which automatically exercises a target system with large volumes of inputs, either \emph{generated} using a model of the input space (e.g., a protocol state machine), or \emph{mutated} from an initial set of regular inputs (seeds). Research on fuzzing is of experimental nature, as this technique lies on heuristics, such as on seed selection and mutation operators. Heuristics leverage empirical insights about typical vulnerabilities, such as, knowing that out-of-bounds integers are often not well managed, mutation operators perform integer operations on the seeds. 

Therefore, one of the key factors for progress in fuzzing research is the availability of \textbf{benchmarks}, that is, representative and well-defined experimental targets for applying heuristic techniques, and for obtaining reproducible and quantitative evaluations. However, there is still a debate about how to benchmark fuzzers, e.g., how to measure progress and depth of fuzzing, how to avoid biases, and whether the targets are representative of real vulnerable systems. Existing proposals for benchmarking fuzzers include \emph{LAVA-M} \cite{dolan2016lava}, a set of open-source program injected with synthetically-generated bugs; \emph{Google FuzzBench} \cite{metzman2020fuzzbench}, a cloud service for running a fuzzer on a set of target programs from the OSS-Fuzz project; the DARPA \emph{Cyber Grand Challenge} corpus \cite{cgc}; and \emph{Magma} \cite{hazimeh2020magma}, a benchmark of targets with manually-curated vulnerabilities. 
However, existing benchmarks focus on \emph{stateless} libraries and programs (e.g., parsers for multimedia file formats, data compression tools, OS utilities, etc.), but have limited or no support for \emph{stateful} software, which would be useful to make progress on fuzzing techniques suitable for protocol testing.

In this paper, we present a new benchmark (\emph{ProFuzzBench}) for stateful fuzzing of network protocols. This benchmark provides a suite of open-source programs that implement popular network protocols, and tools for fully automating the execution of fuzzing experiments. We selected programs representative of the ones adopted in previous research on stateful fuzzing, and of network protocols targeted by commercial and open-source fuzzers. 
Based on our experience with these programs, we discuss some open research issues for stateful network protocol fuzzing, and potential future directions. 
We release the benchmark as open-source, inviting the community to use it for evaluation of new fuzzing techniques, and for further extend it with more targets.

In the following of this paper, Section 2 presents the benchmark targets; Section 3 elaborates on fuzzing automation; Section 4 discusses challenges for applying fuzzers on stateful network protocol implementations; Section 5 concludes the paper.

\section{Benchmark targets}

\begin{table*}[!htp]\centering
\caption{Benchmark targets}\label{tab:benchmark_targets}
\scriptsize
\begin{tabular}{lrrrrrr}\toprule
\textbf{Protocol} &\textbf{Implementation} &\textbf{Description} &\textbf{Protocol states} &\textbf{Related papers} &\textbf{Non-academic fuzzing tools} \\\midrule
RTSP &Live555 &Real-time media streaming &Progress of ongoing streaming &\cite{gaofw,pham2020aflnet} &\cite{beSTORM,Defensics} \\
SMTP &Exim &Email transmission &Message queue, progress of the session &\cite{takanen2018fuzzing} &\cite{beSTORM,Metasploit,Defensics} \\
FTP &ProFTPD, LightFTP &File transfer &CWD, session flags, progress of the session &\cite{takanen2018fuzzing,gascon2015pulsar,pham2020aflnet} &\cite{beSTORM,Metasploit,Defensics} \\
SSH &OpenSSH &Secure remote shell &User authentication progress, session configuration &\cite{zheng2019firm} &\cite{beSTORM,Metasploit,Defensics} \\
TLS &OpenSSL &Secure socket connection &User authentication progress, session configuration &\cite{de2015protocol,walz2017exploiting} &\cite{beSTORM,Defensics} \\
DTLS &TinyDTLS &Secure datagram communication &User authentication progress, session configuration &\cite{fiterau2020analysis} &\cite{Defensics} \\
DNS &dnsmasq &Network domain names &Cached DNS records &\cite{zheng2019firm,takanen2018fuzzing} &\cite{beSTORM,Metasploit,Defensics} \\
SIP &Kamailio &Signaling protocol for real-time sessions &User registrations, progress of the session &\cite{banks2006snooze,alrahem2007interstate,abdelnur2007kif} &\cite{beSTORM,Defensics} \\
DAAP &forked-daapd &HTTP-based audio library streaming &Progress of ongoing streaming, playlist &\cite{shapiro2011identifying} &\cite{beSTORM,Metasploit,Defensics} \\
DICOM &dcmqrscp &Image retrieval &Progress of the session &\cite{wang2019medical} &\cite{Defensics} \\
\bottomrule
\end{tabular}
\end{table*}

We selected a set of network protocols that we wanted to cover with the benchmark. We looked for protocols that are mature and widely used, both by enterprises and by individual users. Moreover, the selection of the protocols took into account which protocols were analyzed by previous fuzzing studies, as reported in academic papers, books, technical blogs, or supported by non-academic fuzzing tools (e.g., commercial products).

For every selected protocol, we looked for open-source software that implements that protocol, to be subject to fuzz testing. In this case too, we looked for mature software suitable for use in real-world applications (thus, worth of being tested for security vulnerabilities), and actively maintained and popular among users. 

\tablename{}~\ref{tab:benchmark_targets} lists the protocols, and the related software implementations, that were selected for the benchmark. In total, we selected 10 protocols, and one software implementation for each of them. FTP is the only exception, for which we included two implementations in the benchmark. Since FTP is often adopted as running example in books and papers to present fuzzing techniques and tools, we found several FTP implementations that were subjected to fuzz testing. Thus, we included more than one software for this protocol.

For each protocol, the table provides a brief description of the objective of the protocol, and of the stateful aspects of the protocol. In many cases, the state of the protocol keeps track of the progress of a communication session between a client and a server. For example, the FTP protocol prescribes initial states for authentication, with messages for providing the username and password; then, authentication is followed by one or more commands for file transfer; and finally, the session ends with a log-out. 
Similarly, in the case of the secure communication protocols (SSH, TLS, DTLS), the protocol state tracks the order of the steps in the authentication process, including the exchange of public key and other information for encryption. 
In some protocols, the protocol state reflects the current state of the server process, and the history of its interactions with the client: for example, in SMTP and DAAP, the server enqueues emails to be transmitted, and songs to be played; while they are processed, the server continues to interact with the client through further messages. 
The benchmark also includes the DNS protocol, despite it is in principle a stateless one. We still opted to include this protocol in the benchmark, since it is a popular fuzzing target, and since its implementations can still provide stateful behavior, such as the caching of records from past queries.

The table provides a non-exhaustive list of papers and books related to each protocol. In these references, the protocols were adopted as case studies for evaluating the effectiveness of new techniques for stateful protocol fuzzing. In the case of some references, the protocols were adopted to provide context for examples, such as to show how to customize and to configure a fuzzing tool for a new protocol. For example, in generational fuzzing, the user needs to provide a model for the protocol under test, either by writing it from scratch, or by reusing an existing protocol model (e.g., based on protocol specifications).

Finally, the table lists non-academic fuzzing tools (commercial and open-source) that provide support for each protocol. The commercial Defensics tool from Synopsis (originally developed by Codenomicon) and beSTORM from Beyond Security, support an extensive set of network protocols, including the ones that are part of the benchmark. The Metasploit Framework (best known for its exploit database) also provides a small set of fuzzers for few basic network protocols, which were included in the benchmark. The DAAP protocol is not directly supported by these fuzzers, but can be fuzzed using these fuzzing tools using their support for HTTP.

\section{Automation}

\begin{figure*}
  \includegraphics[width=\textwidth]{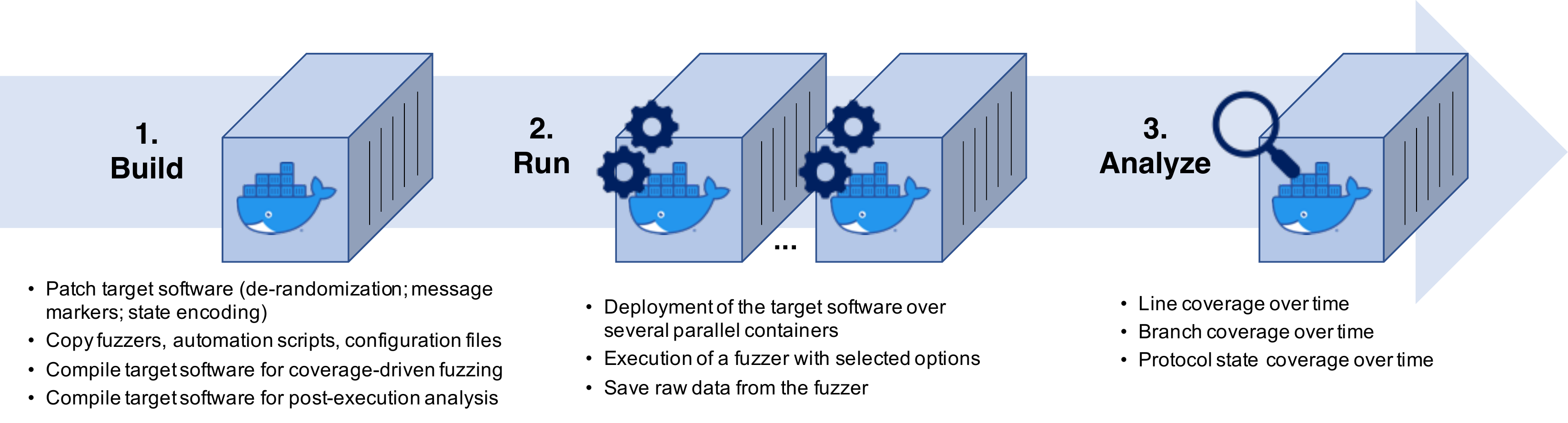}
  \caption{Workflow of benchmark automation.}
  \label{fig:workflow}
\end{figure*}

The protocol fuzzing benchmark has been published as a set of utilities for automating the compilation, configuration, execution, and analysis of target software (Figure~\ref{fig:workflow}). 
These utilities are based on Docker to automate the deployment of the target software and of fuzzing tools, in order to achieve reproducible experiments, and to support the comparative analysis of different fuzzing techniques under controlled conditions. 

The first stage is the build of a container image. The target software, and a set of protocol fuzzing tools, are first cloned from their open-source repositories. The tools currently included in the benchmark are \textsc{AFLnwe} (a basic variant of AFL, to support fuzzing over network sockets) and \textsc{AFLnet} (a protocol-aware fuzzer, also based on AFL) \cite{pham2020aflnet}. The benchmark comes with a set of patches for each target software, which are applied during this state. 
For most target software, the patches de-randomize the program, such as, by embedding a fixed seed in the source code. De-randomization is important to have reproducible behavior, i.e., if the program is executed again with the same input, then the same execution path is covered. This reproducible behavior is an implicit assumption for the correct application of coverage-driven fuzzing techniques. Moreover, de-randomization supports the comparison of different fuzzing techniques, as it makes it less likely that a fuzzer perform better than other by chance, rather than for its merits. 
In some cases, the patches decrease or remove delays in the program (e.g., uses of \texttt{sleep()}), which are adopted for synchronization in network settings, but which unnecessarily slow down fuzzing since it can be performed locally with smaller delays. 
Other patches make the program compatible with fuzzers. For example, AFL uses UNIX signals to manage the target process and to detect its failures. Thus, any custom handler for these signals should be disabled to avoid interfering with AFL-based fuzzing tools.

Some patches of the benchmark support the experimentation of tools tailored for protocol fuzzing. For example, \textsc{AFLnet} \cite{pham2020aflnet} takes in input a corpus of network traffic traces, where an individual fuzz input is represented by one traces, which consists of a sequence of messages exchanged with the target software. In order to test more states of the protocol, a tool may need to focus fuzzing on a specific message in the sequence. Thus, \textsc{AFLnet} performs a parsing of the traffic traces, by identifying the position of the individual messages. \textsc{AFLnet} includes several parsers for the supported protocols. Where available, the parsers look for marker characters to split the messages, such as, the characters \texttt{0x0D0A} in FTP. Some protocol implementations do not provide such markers (e.g., a message ends when the connection is closed, without an explicit marker). In these cases, the patches introduce such characters to simplify the implementation of message parsers.

Moreover, the patches support protocol fuzzing tools that keep track of protocol states in order to maximize state coverage. Some protocols, such as FTP, provide distinctive status codes in response messages from the server to the client (e.g., the number $230$ when the login succeeded and the server awaits for commands or number $257$ when a new folder has been successfully created). However, it might not be the case for all protocols. For instance, RTSP protocol uses the same response code $200$ for all successful requests, regardless of the current server state. To make stateful fuzzing more effective, our benchmark patches the target (Live555 server) to send responses with different status codes e.g., $201$ when a PLAY command has been processed and $203$ for a PAUSE command.

The (patched) target software is then compiled within the container image. It is compiled a first time using the AFL compiler wrapper (i.e., afl-clang-fast/afl-clang-fast++), in order to include the runtime instrumentation needed by AFL-based coverage-driven fuzzing tools. This binary will be the target for fuzzing. The target software is compiled a second time with support for \texttt{gcov}: this binary will be used in the last stage of the benchmark, in order to compute coverage metrics after fuzzing. The build stage of the benchmark finalizes the container image by adding target-specific files (e.g., configuration file, media sources). Finally, the fuzzing tools and automation scripts are copied in the container image.

In the second stage, the benchmark runs the selected fuzzer on the target software, using the container image. In order to compare different fuzzers in a rigorous way, the same fuzzing experiment should be repeated several times. In this way, two fuzzers can be compared with respect to their results over several repetitions, using statistical techniques analysis to achieve confidence that one of the fuzzers is indeed superior to the other \cite{klees2018evaluating}. Therefore, the utilities provided with the benchmark allow the user to run multiple instances of the same container in parallel, by taking advantage of multicore architectures. Finally, raw data from the fuzzer is copied from the container to  the host machine. The raw data include inputs generated by the fuzzer that covered a new path in the program (e.g., ``interesting inputs'' in AFL); inputs that discovered new state transitions in the protocol state machine (e.g., as in \textsc{AFLnet}); and inputs that caused a crash or hang of the target software.

The final stage analyzes the raw data to generate statistics about the fuzzing experiments. It runs again the inputs using the target software compiled with \texttt{gcov}, in order to obtain line- and branch-level information about the coverage of each input. Then, it generates time series in CSV format for the line and branch coverage, based on the timestamps at which the inputs were generated, to be used for plotting and for comparing the fuzzers.

\begin{figure}
  \includegraphics[width=\columnwidth]{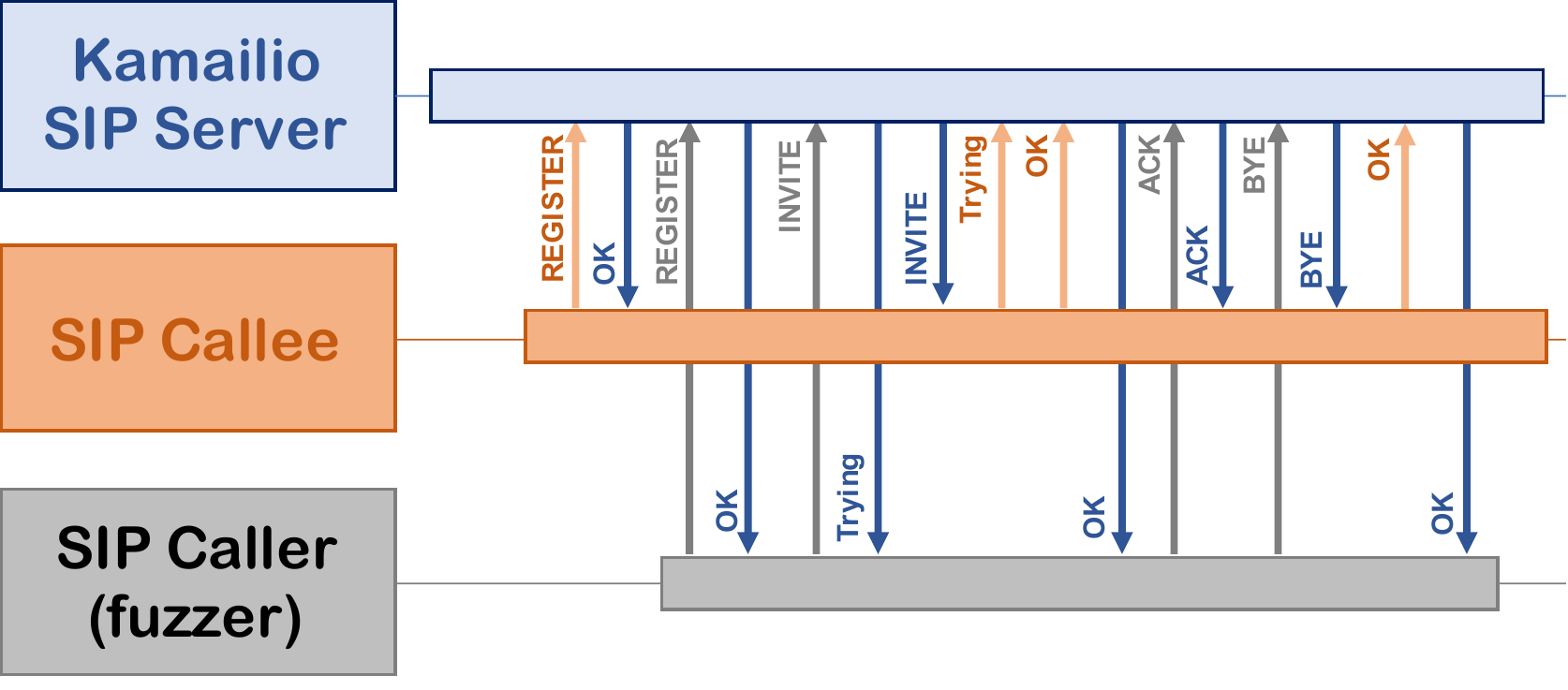}
  \caption{SIP protocol fuzzing.}
  \label{fig:sip-protocol}
\end{figure}

\section{Discussion}

The availability of this benchmark is an opportunity for research in new techniques for protocol fuzzing. In selecting the target software, and in integrating them with a fuzzer, we found the following open research issues.

\textbf{Configuration and multi-party protocols.} 
The fuzzing of network protocols requires the configuration of their software implementations. Selecting a specific configuration determines which parts of a network server can be tested by fuzzing. The configuration space can be quite complex for real-world network servers, but the problem has not been studied yet. Previous research has been mostly focused on stateless programs, such as libraries and command-line utilities,  where the main challenge comes from the complexity of the inputs and of their constraints for reaching deep execution paths. As the configuration affects the attack surface, it indirectly becomes a concern for fuzzing. 

In particular, protocols can require to run more than two parties. Existing fuzzing techniques only adopt a client-server scheme, where the fuzzer acts as client. However, in the case of SIP protocol, at least three parties are required: the server under test, a ``caller'' and a ``callee''. One or more ``callee'' and ``caller'' must first register on the server; then, a caller can contact the server, by indicating the name of a callee; the server will act as proxy between the callee and the caller, in order to establish a communication (e.g., a VoIP call) between them. In the process, the server will handle the ``ringing'', the possibility that the callee is busy, etc.

Figure~\ref{fig:sip-protocol} shows an example of interactions between the fuzzing tool and the Kamailio SIP target software. To fuzz a SIP session, which establishes a call between a caller and callee, the fuzzer should take one of these roles (in our benchmark, it is the caller). In addition to the fuzzer, it is necessary to have another process to take the other role. Therefore, the benchmark comes with scripts to automate the execution both of the SIP server and of a callee using a SIP client; after the callee has registered, the fuzzers can replay and fuzz the sequences of message from the corpus. The corpus has been prepared to match the configuration of the processes, i.e., the same port numbers and client names are both referenced in corpus, and used by the SIP client and by the fuzzer.

In more complex protocols, such as for IP routing and network management, even more parties may be required for a meaningful testing. When setting up multiple parties, identifying a configuration for the target becomes even more challenging. Moreover, these protocols are beyond the support of current fuzzers, as they focus on fuzzing an individual stream of data to the server.

\textbf{Deterministic execution.} 
Another challenge towards fuzzing a complete network server is non-deterministic behavior, due to the extensive use of threads and event-based I/O. When the same input is applied to the target software, its non-determinism causes variations of the coverage of program paths, which in turn hinders coverage-driven fuzzing. For this reason, the AFL fuzzer performs a ``calibration'' on every new input that expands the coverage, in order to assess whether the execution is deterministic, and to report the percentage of anomalous new inputs (the \emph{stability} metrics). In the benchmark, we followed the best practices suggested for AFL \cite{afldoc}, by disabling threads where possible; or, by replacing kernel-level threads (as in \emph{pthreads}) with user-level ones (as in \emph{GNU Pth}) and related I/O APIs, as we did in the \emph{forked-daapd} target. Despite these countermeasures, the reported stability is still lower than 50\% in several benchmarks, due to the use of threads and I/O in libraries called by the target software. Moreover, user-level threads increase the complexity of the testing setup (e.g., in \emph{forked-daapd}, it hinders the saving of coverage data).

\textbf{Test execution speed.} 
The execution speed of fuzz testing for network servers can be quite slow. Compared to fuzzing libraries and command-line utilities (which can reach up thousands of executions per second in AFL), fuzzing network servers is slowed down by the complex initialization of the process (e.g., to initialize a cache or a database), the networking overhead (due to communication and context switches between processes), and the long sequences of messages in a test case. For example, in fuzzing SIP, the initialization was slowed down by the synchronization between the fuzzer and the other processes (the server and the SIP callee), with less than 5 executions per second. Notably, in fuzzing DAAP, the server under test takes about 2 seconds to reliably complete its initialization hence the fuzzing speed significantly drops.

\textbf{State identification}
The goal of protocol fuzzing is to generate tests that cover protocol states. One important open area of research is to perform fuzzing when a formal specification of the protocol state machine is not available (e.g., using model inference techniques, which reverse a protocol by trial-and-error). In our benchmark, we also faced the problem of identifying the current state of a protocol at run-time, when the protocol does not explicitly encode the current state in the messages (e.g., RTSP protocol).

\section{Conclusion}

In this paper, we present a new benchmark (\emph{ProFuzzBench}) for stateful fuzzing of network protocols. The benchmark currently supports 10 protocols and it can be extended. Based on our preliminary experiments, we also discuss open challenges for future research. As future work we plan to add more targets and integrate more fuzzers into the benchmark. Moreover, we will run large-scale experiments on the benchmark and analyze the results to gain more insights into the identified challenges.

\section*{Benchmark availability}

The source code of \emph{ProFuzzBench} is available online on GitHub at: \url{https://github.com/profuzzbench/profuzzbench}


\bibliographystyle{ACM-Reference-Format}
\bibliography{sample-base,references}



\end{document}